\documentclass{article}
\usepackage{spconf,amsmath,graphicx,amssymb,xcolor}
\usepackage{algpseudocode}
\usepackage{algorithm}
\usepackage{amsfonts}
\usepackage{booktabs}
\usepackage{soul}

\def\N{{\mathcal{N}}}
\def\ba{{\Bar{\alpha}}}

\title{SDEMG: Score-Based Diffusion Model for Surface Electromyographic Signal Denoising}
\name{Yu-Tung Liu$^{1,5}$ \qquad Kuan-Chen Wang$^{2,5}$ \qquad Kai-Chun Liu$^{3}$ \qquad Sheng-Yu Peng$^{4}$\qquad Yu Tsao$^{5}$}
\address{
\small{
$^{1}$Department of Electronics and Electrical Engineering, National Yang Ming Chiao Tung University, Taiwan} \\ 
\small{$^{2}$Graduate Institute of Communication Engineering, National Taiwan University, Taiwan} \\
    \small{$^{3}$Department of Electronic Engineering, National Taipei University of Technology, Taiwan}\\ 
    \small{$^{4}$Department of Electrical Engineering, National Taiwan University of Science and Technology, Taiwan}\\
    \small{$^{5}$Research Center for Information Technology Innovation, Academia Sinica, Taiwan}
    \\
\normalsize{tonyliu.ee09@nycu.edu.tw, d12942016@ntu.edu.tw, kaichunliu@ntut.edu.tw, sypeng@mail.ntust.edu.tw, yu.tsao@citi.sinica.edu.tw}
    }
\begin{document}
\ninept
\maketitle
\begin{abstract}
Surface electromyography (sEMG) recordings can be influenced by electrocardiogram (ECG) signals when the muscle being monitored is close to the heart. Several existing methods use signal-processing-based approaches, such as high-pass filter and template subtraction, while some derive mapping functions to restore clean sEMG signals from noisy sEMG (sEMG with ECG interference). Recently, the score-based diffusion model, a renowned generative model, has been introduced to generate high-quality and accurate samples with noisy input data. In this study, we proposed a novel approach, termed SDEMG, as a score-based diffusion model for sEMG signal denoising. To evaluate the proposed SDEMG approach, we conduct experiments to reduce noise in sEMG signals, employing data from an openly accessible source, the Non-Invasive Adaptive Prosthetics database, along with ECG signals from the MIT-BIH Normal Sinus Rhythm Database. The experiment result indicates that SDEMG outperformed comparative methods and produced high-quality sEMG samples. The source code of SDEMG the framework is available at: https://github.com/tonyliu0910/SDEMG

\end{abstract}
\begin{keywords}
Surface electromyography, Score-based diffusion model, ECG interference removal, Deep neural network
\end{keywords}

\section{Introduction}

\label{sec:intro}
Surface electromyography (sEMG) records the biopotential generated by motor units during muscle contractions. This technique can noninvasively provide valuable insights into muscle anatomy and physiology, and thus sEMG finds applications in various clinical areas, including neuromuscular system investigation~\cite{tang2018novel}, rehabilitation~\cite{engdahl2015surveying}, stress monitoring~\cite{wijsman2013wearable}, assessment of neuromuscular or respiratory disorders~\cite{domnik2020clinical, vandenbussche2015assessment}, and prosthesis control~\cite{ma2014hand}. In these applications, sEMG recordings may be subject to electrocardiogram (ECG) interference if the measurement is taken near the heart~\cite{xu2020comparative, guo2023morphological}. The ECG contamination can distort the amplitude and frequency aspects of sEMG signals, posing a challenge in extracting meaningful information. Hence, developing effective ECG removal methods is crucial to enhance signal quality for various clinical and human-computer interaction applications.

sEMG and ECG signals have frequency bands between 10 to 500 Hz and 0 to 100 Hz, respectively~\cite{winter2009biomechanics}. The overlapping frequency bands pose difficulties in segregating the two signals. To address this issue, several single-channel ECG removal methods have been developed, such as high-pass filters (HP) and template subtraction (TS)~\cite{xu2020comparative,drake2006elimination}. However, HP causes distortion by removing the low-frequency part of sEMG signals, and TS relies on the assumption that ECG is quasi-periodic and sEMG follows a zero-mean Gaussian distribution, which may not hold in real-world scenarios. These limitations make these ECG removal methods struggle under demanding conditions, such as low signal-to-noise (SNR) ratios.
Recently, neural networks (NNs) have been widely adopted for their powerful nonlinear mapping capabilities in signal enhancement~\cite{lu2013speech,chiang2019noise}. Some studies have applied NNs for sEMG denoising~\cite{kale2009intelligent,wang2023ecg}. In ~\cite{wang2023ecg}, Wang et al. developed an NN-based ECG removal method employing a fully convolutional network (FCN) as a denoising autoencoder (DAE). The experimental results in ~\cite{wang2023ecg} show that FCN outperformed traditional methods in ECG artifact removal, while it also introduced some signal distortion, which can be problematic in clinical applications. Therefore, it is desirable to develop more effective ECG removal methods.


The score-based diffusion model, a generative model, has excelled in producing high-quality data in various tasks, including image and acoustic generation~\cite{ho2020denoising,song2020score,chen2020wavegrad}; various studies have also incorporated the model into signal enhancement~\cite{lu2021study,lu2022conditional,li2023descod}. Compared to discriminative NNs, the score-based diffusion model could observe and model the data distribution and reconstruct the desired sample through sampling steps. Also, the diffusion process and the reverse process are relatively more tractable and more flexible compared to other generative models, such as generative adversarial networks (GANs) and variational autoencoders (VAEs), i.e., we can determine the sample quality by adjusting the hyperparameters empirically. The features of the score-based diffusion model mentioned above allow us to model the distribution of the dataset as a whole and generate samples of higher quality based on the given conditions.

To refine sEMG denoising performance, this study proposes SDEMG, a conditional score-based diffusion model for sEMG denoising. The proposed method progressively adds isometric Gaussian noise to the clean sEMG during the diffusion process. In the reverse process, we leverage the sEMG waveform contaminated by ECG and the noise scale variable as conditions for the NN, and the Gaussian noise is reverted to the clean sEMG segment. Experimental results show that SDEMG outperforms the previous FCN-based denoising method in signal quality, providing a refined ECG removal approach for clinical sEMG applications.

The remainder of this paper is organized as follows. Section 2 reviews related works. Section 3 introduces the proposed approach. Section 4 presents the experimental setup and results. Finally, Section 5 concludes the paper and discusses future works.

\section{Related Work}
\label{sec:related}

\subsection{ECG interference removal methods}
\label{ssec:previous}
Several single-channel ECG removal methods have been developed in previous studies, including HP and TS~\cite{xu2020comparative,drake2006elimination}. HP removes the frequency band of the ECG, inevitably leading to the loss of the low-frequency part of the sEMG signal. In contrast, TS removes ECG artifacts in the time domain. It extracts ECG templates for subtraction by either filtering or waveform averaging~\cite{xu2020comparative,junior2019template}, and the ECG artifacts are subtracted from the contaminated sEMG waveform. However, the effectiveness of TS relies on the assumption that sEMG signals are zero-mean Gaussian distributions, which may not be satisfied in real-world scenarios. This study applies both HP and TS to serve as comparative sEMG denoising methods. HP is implemented with a cutoff frequency of 40 Hz, and the TS method was followed using the HP method for optimal results.

Beyond conventional techniques, NN-based methods have demonstrated exceptional performance in sEMG denoising~\cite{kale2009intelligent,wang2023ecg}. In~\cite{wang2023ecg}, Wang et al. proposed an FCN as a DAE to eliminate ECG interference from sEMG. The proposed FCN consists of two parts: an encoder and a decoder. Both parts consist of multiple convolutional layers with different filter sizes and strides. The encoder receives contaminated sEMG waveform as input and performs downsampling to generate high-level feature maps. Conversely, the decoder attempts to reconstruct the clean sEMG waveform from the encoder output. According to the study, the FCN demonstrates more effective denoising performance than conventional methods.

\subsection{Score-based diffusion models}
\label{ssec:diffusion}
Score-based diffusion models are a category of deep generative models initially devised for image generation, and they have further been adapted to signal enhancement, including speech enhancement and ECG denoising~\cite{lu2022conditional,li2023descod}. It is preferable to other generative methods, such as GANs and VAEs. To elaborate, GANs are notoriously unstable during the training process, and VAEs lack sample quality. Score-based diffusion models overcome these limitations and achieve remarkable performance in numerous applications~\cite{ho2020denoising,kong2020diffwave, huang2023noise2music}.

The process of the score-based diffusion model starts with gradually adding Gaussian noise to input data at different scales. The NN within the diffusion model aims to estimate the noise segment at each particular time step. Subsequently, desired samples are generated during the sampling process.
\begin{equation}
\label{eq: Stein score function}
    s(y)=\nabla_{x} \log p(x).
\end{equation}
To further introduce the score-based diffusion model, we start with modeling the underlying data distribution with the Stein score function as shown in Eq. (\ref{eq: Stein score function}), which is the gradient of data log-density $\log p(x)$ concerning data $x$:
\begin{equation} 
\label{eq: Fisher divergence}
\mathbb{E}_{p(\mathbf{x})}[\| \nabla_\mathbf{x} \log p(\mathbf{x}) - \mathbf{s}_\theta(\mathbf{x}) \|_2^2].
\end{equation}
We can train an NN, a score-based model $s_{\theta}$, to minimize the Fisher divergence between the model and the actual data distribution as shown in Eq. (\ref{eq: Fisher divergence}). As presented, we calculate the L2 distance between the target data score and the score-based model with score-matching methods~\cite{hyvarinen2005estimation, song2020sliced}:
\begin{equation}
\mathbf{x}_{i+1} \gets \mathbf{x}_i + \epsilon \nabla_\mathbf{x} \log p(\mathbf{x}) + \sqrt{2\epsilon}~ \mathbf{z}_i, \quad i=0,1,\cdots, K. 
\label{eq:langevin} 
\end{equation}
After training the score-based model $s_{\theta}$, we can generate samples from the specific density, $\Tilde{x} \sim p(x)$, via Langevin dynamics presented in Eq. (\ref{eq:langevin}):

\begin{figure}[t!]
    \centering
    \includegraphics[width=\linewidth]{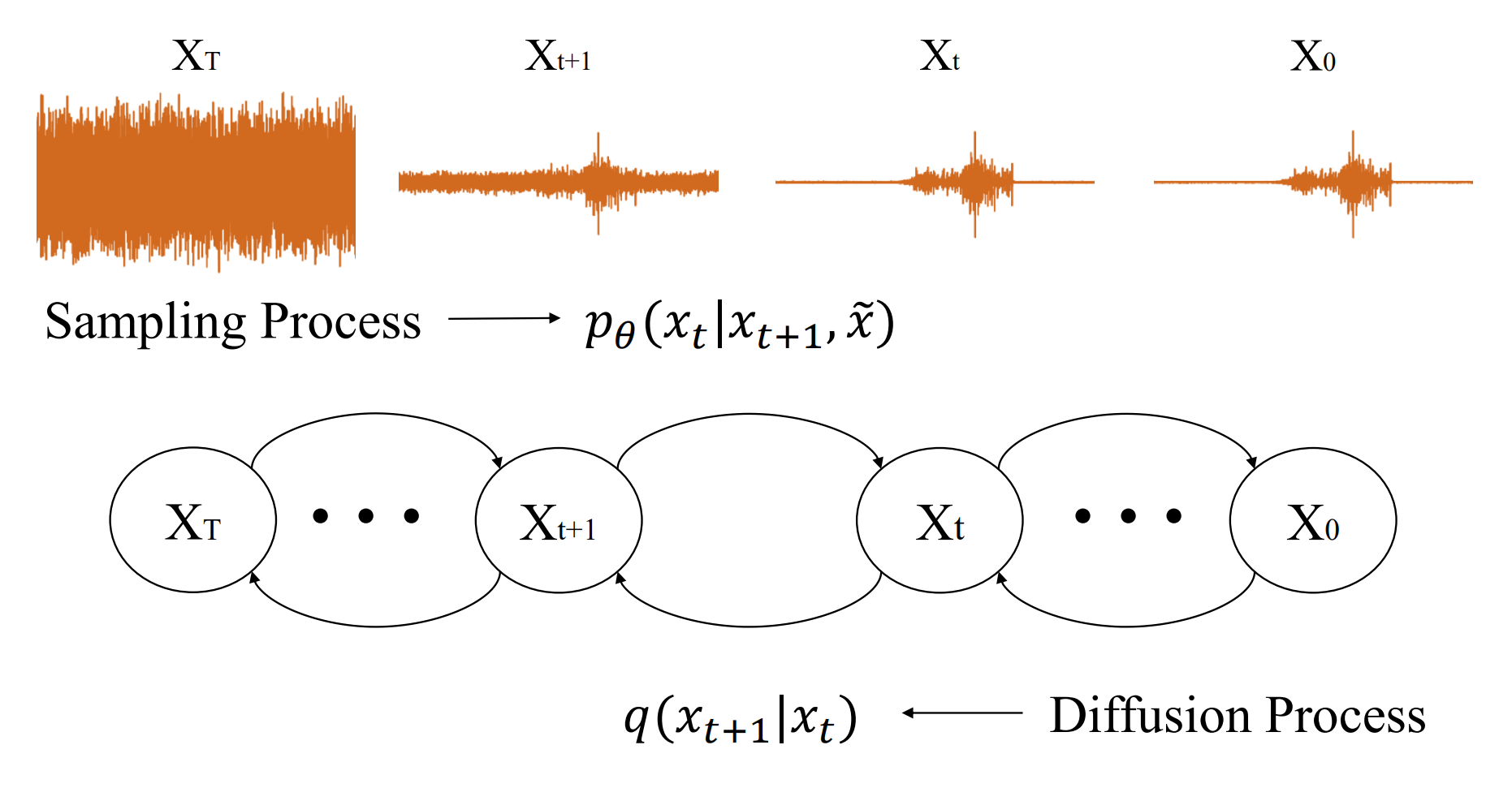}
    \caption{The diffusion process and the sampling in SDEMG. }
    \label{fig: denoise step}
\end{figure}

\section{The Proposed SDEMG Method}
\label{sec:proposed}
\subsection{Training process}
\label{subsec: training}
In this study, we trained a score-based model to estimate the noise distribution and generate high-quality and high-fidelity sEMG samples based on the diffusion probabilistic model proposed by Ho et al.~\cite{ho2020denoising}. The original model was devised for unconditional image generation, i.e., the model generates samples from Gaussian noise without any inputs. In our case, we aim to retrieve clean sEMG samples from noisy sEMG waveforms contaminated with ECG. Therefore, we require the noisy signal $\Tilde{x}$ as a condition. The score-based denoising framework takes sEMG mixed with actual noise as input and attempts to reconstruct the clean waveform. 

The score-based diffusion framework comprises two steps: a diffusion process (algorithm~\ref{algorithm: diffusion}) and a sampling process (algorithm~\ref{algorithm: sampling}). The diffusion process (algorithm~\ref{algorithm: diffusion}), otherwise forward process, is defined as a Markov Chain as shown in Eq. (\ref{eq: markov}): 

\begin{equation}
    \label{eq: markov}
    q(x_{1:T}|x_{0})=\prod_{t=1}^{T}q(x_{t}|x_{t-1}).
\end{equation}

During this process, Gaussian noise is gradually added (Eq. (\ref{eq: iteration})) with respect to different noise schedules $\beta_{1},\dots,\beta_{N}$. The original signals will be perturbed more significantly as $t$ becomes larger. 

\begin{equation}
\label{eq: iteration}
   q(x_{t}|x_{t-1})=\mathcal{N}(x_{t};\sqrt{1-\beta_{t}}x_{t-1},\beta_{t}I).  
\end{equation}

Ho et al.~\cite{ho2020denoising} observed that: Let $\alpha_{t} = 1-\beta_{t}$ and $\bar{\alpha}_t = \prod_{i=1}^t \alpha_i$, the resultant distribution of the diffusion process can be presented in Eq. (\ref{eq: reparameterization}) with variance $1-\bar{\alpha_{t}}$:

\begin{equation}
\label{eq: reparameterization}
q(\mathbf{x}_t \vert \mathbf{x}_0) = \mathcal{N}(\mathbf{x}_t; \sqrt{\bar{\alpha}_t} \mathbf{x}_0, (1 - \bar{\alpha}_t)\mathbf{I}).
\end{equation}

As mentioned earlier, to avoid signal distortion and generate high-quality samples, during the training process, we introduce noisy sEMG as the condition to assist the model in retrieving more features regarding the data distribution. Furthermore, Song \& Ermon~\cite{song2019generative} discovered that the selection of noise schedule is critical to sample quality, and Chen et al.~\cite{chen2020wavegrad} proposed the continuous condition $\bar{\alpha}$ instead of discrete time step variable $t$ will provide more auxiliary information to the model. In the methods proposed by Li et al.~\cite{li2023descod}, the reparameterized condition was also adopted and generated high-quality samples.  Consequently, we utilized the cosine beta scheduler as it was more helpful for the model to generate more robust predictions~\cite{nichol2021improved} and adapted similar condition methods to our denoising framework and selected the corresponding noise scale $\bar{\alpha}$ from the noise scale schedule $\gamma = \{1, \sqrt{\Bar{\alpha_{0}}}, \dots, \sqrt{\Bar{\alpha_{T}}}\}$ as one of the inputs. As shown in Algorithm ~\ref{algorithm: diffusion}, the model will receive clean sEMG $x_{0}$, noisy sEMG $\tilde{x}$, and noise scale parameter $\bar{\alpha}$ as inputs and aims to predict the isotropic Gaussian noise $\epsilon$. We will minimize the mean squared error of the predictions during the training process. 

\begin{algorithm}[t]
\begin{algorithmic}[1]
\caption{Training}
\label{algorithm: diffusion}
\Repeat
    \State $x_{0}, \Tilde{x} \sim q(x_{0}, \Tilde{x})$
    \State $t \sim Uniform(\{1,\dots,T\})$
    \State $\ba \sim Uniform(\gamma_{t-1}, \gamma)$
    \State $\epsilon \sim \N(0, I)$
    \State \text{Take gradient descent step on}
        \[\nabla_{\theta}||\epsilon-\epsilon_{\theta}(\sqrt{\ba}x_{0}+\sqrt{1-\ba}\epsilon, \Tilde{x}, \ba)||^{2}\] 
\Until{Converged}
\end{algorithmic}
\end{algorithm}

\subsection{Sampling process}
\label{subsec: sampling}
In the reverse sampling process, as shown in Algorithm~\ref{algorithm: sampling}, the initial sample is random Gaussian noise $x_{T}$. The denoising framework will reconstruct the desired sEMG signals through $T$ sampling steps. The sampling process is defined as a reverse Markov chain in Eq. (\ref{eq: reverse markov}):
\begin{equation}
\label{eq: reverse markov}
p_\theta(\mathbf{x}_{0:T}) = p(\mathbf{x}_T) \prod^T_{t=1} p_\theta(\mathbf{x}_{t-1} \vert \mathbf{x}_t, \tilde{x}).
\end{equation}
During each iteration, SDEMG will generate a sample $x_{t-1}$ from the previous sample $x_{t}$ and the input noisy data $\tilde{x}$:
\begin{equation}
\label{eq: reverse iteration}
p_\theta(\mathbf{x}_{t-1} \vert \mathbf{x}_t, \tilde{x}) = \mathcal{N}(\mathbf{x}_{t-1}; \boldsymbol{\mu}_\theta(\mathbf{x}_t, t, \tilde{x}), \boldsymbol{\Sigma}_\theta(\mathbf{x}_t, t, \tilde{x})).
\end{equation}

\subsection{Model Architecture}
\label{subsec: model}
The network within SDEMG is adapted from the model proposed by Li et al.~\cite{li2023descod}. The model comprises two data streams, one for the clean signal and the other for the noisy data, and the bridges to aggregate the features extracted from different layers. The data streams are composed of multiple Half Normalized Filters (HNF), which are computation blocks adopted from Multi-Kernel Filter design~\cite{romero2021deepfilter} and HiNet~\cite{chen2021hinet}. The HNF blocks act as a feed-forward module that combines features extracted by different convolutional layers with various receptive fields with channel-wise concatenation. Moreover, half of the features will be normalized as this technique has demonstrated more stability during the training process~\cite{chen2021hinet}. These blocks utilize residual connections to produce outputs from input data and processed features. The Bridge blocks are modified from the feature-wise linear modulation (FiLM) block~\cite{dumoulin2018feature} and condition on the input noise scale $\sqrt{\bar\alpha}$. Notably, we increased the input dimension of the model from 64 to 128 to better accommodate the complexity and dynamic nature of sEMG data, since sEMG is inherently more intricate and variable than ECG data.

\section{Experiments}
\subsection{Datasets}
The sEMG signals used in this study are from the DB2 subset of the Non-Invasive Adaptive Prosthetics (NINAPro) database~\cite{atzori2014electromyography}, which contains 12 channels of sEMG recordings of hand movement made by 40 intact subjects. Notably, the sEMG recordings were acquired from the upper limb. The DB2 subset includes 3 sessions, Exercise 1, 2, and 3, which involve 17, 22, and 10 movements, respectively. Each movement was repeated six times for five seconds, followed by a three-second rest interval. Previous studies~\cite{wang2023ecg,machado2021deep} have applied filters to obtain clean sEMG data from this database.

As for ECG interference, this study adopts MIT-BIH Normal Sinus Rhythm Database (NSRD) from the PhysioNet database~\cite{goldberger2000physiobank}. In this data set, 2-channel ECG recordings were collected from 18 healthy subjects, with a sampling rate of 128 Hz. Previous studies have adopted the ECG signals from this data set as the interference in sEMG signals ~\cite{wang2023ecg, machado2021deep}.

\begin{algorithm}[t]
\begin{algorithmic}[1]
\caption{Sampling}
\label{algorithm: sampling}
\State $x_{T} \sim \N(0,I)$
\For{$t=T, \dots, 1$}
\State$z \sim \N(0, I)$ \textbf{if} $z>1$ \textbf{else} $z=0$ 
\State $x_{t-1}=\frac{1}{\alpha_{t}}\left(x_{t}-\frac{1-\alpha_{t}}{\sqrt{1-\ba_{t}}}\epsilon_{\theta}(x_{t},\Tilde{x},\sqrt{\Bar{\alpha_{t}}})\right)+\sigma_{t}z$
\EndFor
\State \Return $x_{0}$
\end{algorithmic}
\end{algorithm}

\subsection{Data preprocessing and preparation}
The sEMG data were processed by a 4th-order Butterworth bandpass filter with cutoff frequencies of 20 and 500 Hz and downsampled to 1kHz. Subsequently, all sEMG signals were normalized and divided into 10-second segments. ECG data in Channel 1 from the MIT-BIH NSRD were filtered by a 3rd-order Butterworth high-pass and low-pass filter with cutoff frequencies of 10 and 200 Hz to discard possible noise in ECG signals~\cite{xu2020comparative}.

sEMG segments of Channel 2, Exercise 1 and Channel 2, Exercise 3 from 30 subjects were selected as the training and validation set, respectively. For each segment in the training set, 10 randomly selected ECG signals from 12 subjects in the MIT-BIH NSRD were considered ECG artifacts and were superimposed onto the clean sEMG segments at 6 SNRs (-5, -7, -9, -11, -13, and -15 dB). For the validation set, three other subjects in the MIT-BIH NSRD were considered ECG artifacts, and the SNRs are identical to the training sets. 

The mismatch conditions between the training and testing sets were considered to evaluate the generalizability of the proposed framework. sEMG segments of Channel 9, 10, 11, and 12, Exercise 2, from the remaining 10 subjects were selected as the testing set. The ECG data from the remaining 3 subjects (16420, 16539, and 16786) were selected for interference, and the SNRs are -14-0 dB with an increment of 2 dB. Please note that sEMG subjects, sEMG movements, sEMG channel, ECG subjects, and SNRs are entirely different from the setting of the training set.

\subsection{Evaluation metrics}
\label{ssec:metrics}
To compare SDEMG with prior methods, this study follows the earlier work~\cite{wang2023ecg,chiang2019noise,zhang2021eegdenoisenet} and evaluates the performance with two aspects of criteria: signal reconstruction quality and feature extraction errors. SNR improvement (SNR$_{imp}$) and root-mean-square error (RMSE), respectively, reflect the quality of the reconstructed signal by presenting the difference of SNR between the inputs and the outputs and the variance between the outputs and the ground truth. Furthermore, the RMSE of the average rectified value (ARV) and mean frequency (MF) feature vectors are adopted as the metrics to evaluate the features extracted from the sEMG signals~\cite{xu2020comparative}. The settings for calculating these metrics follow the previous study~\cite{wang2023ecg}. Better signal reconstruction quality and higher fidelity can be represented by smaller RMSE values, larger SNR$_{imp}$ values, and smaller RMSE values of the extracted feature vectors ARV and MF.

\begin{table}[t!]
\centering
\caption{Overall performance of HP, TS, FCN, and SDEMG.}
\smallskip
\label{tab: result-table}
\resizebox{1\linewidth}{!}{
\begin{tabular}{@{}clclc@{}}
\toprule
            & \multicolumn{1}{c}{SNR$_{imp}$} (dB)& RMSE        & \multicolumn{1}{c}{RMSE$_{ARV}$} & RMSE$_{MF}$  (Hz)   \\ \midrule
HP          & 13.885                   & 1.735e-2              & 3.06e-3             & 17.688               \\
TS          & 14.279                   & 1.626e-2              & 3.86e-3             & 23.149               \\
FCN         & 17.758                   & 1.178e-2              & 3.86e-3             & 18.038               \\ 
\textbf{SDEMG(Ours)} & \textbf{18.467} & \textbf{1.138e-2}   & \textbf{2.81e-3}    & \textbf{14.435}       \\ \midrule
\end{tabular}
}
\end{table}

\begin{figure}[t!]
    \centering
    \includegraphics[width=\linewidth]{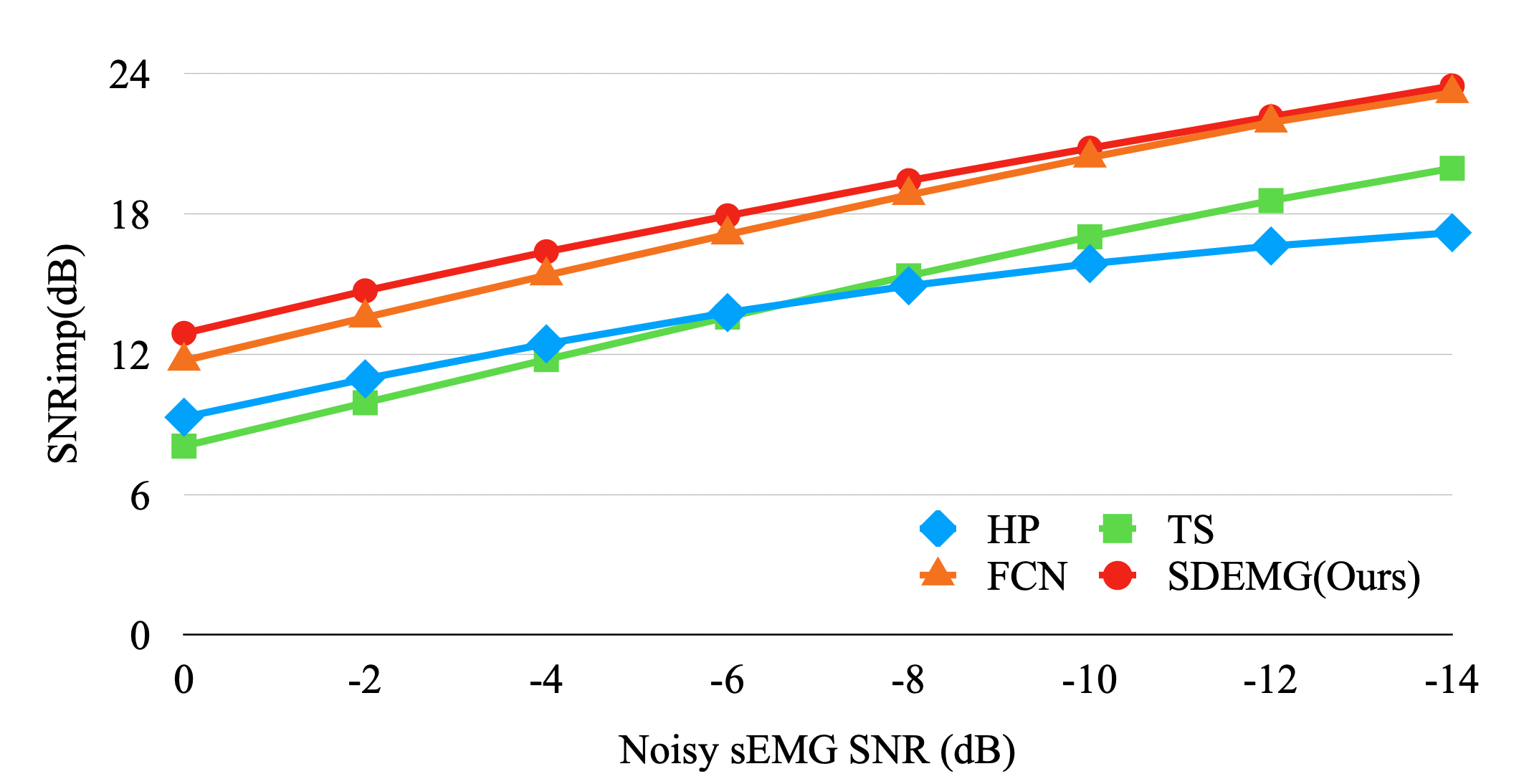}
    \caption{Comparison of all methods with SNR$_{imp}$ under different SNR input.}
    \label{fig: snr_imp}
\end{figure}


\subsection{Results and discussion}
\label{ssec:result}
We compared the performance of SDEMG with three previous methods, namely HP, TS, and FCN~\cite{wang2023ecg}. Table~\ref{tab: result-table} illustrates the overall performance of all methods evaluated by SNR$_{imp}$, RMSE, and RMSE of ARV and MF. The results show that SDEMG can achieve higher SNR improvement, lower RMSE value, and better ARV and MF. Furthermore, Fig.~\ref{fig: snr_imp} presents the SNR$_{imp}$ of ECG elimination approaches under various SNRs. The proposed approach performs best across all SNR conditions, demonstrating its superior effectiveness and robustness in ECG removal.

\begin{figure}[t!]
\begin{minipage}[b]{.48\linewidth}
  \centering
  \centerline{(a)}
  \centerline{\includegraphics[width=4.0cm]{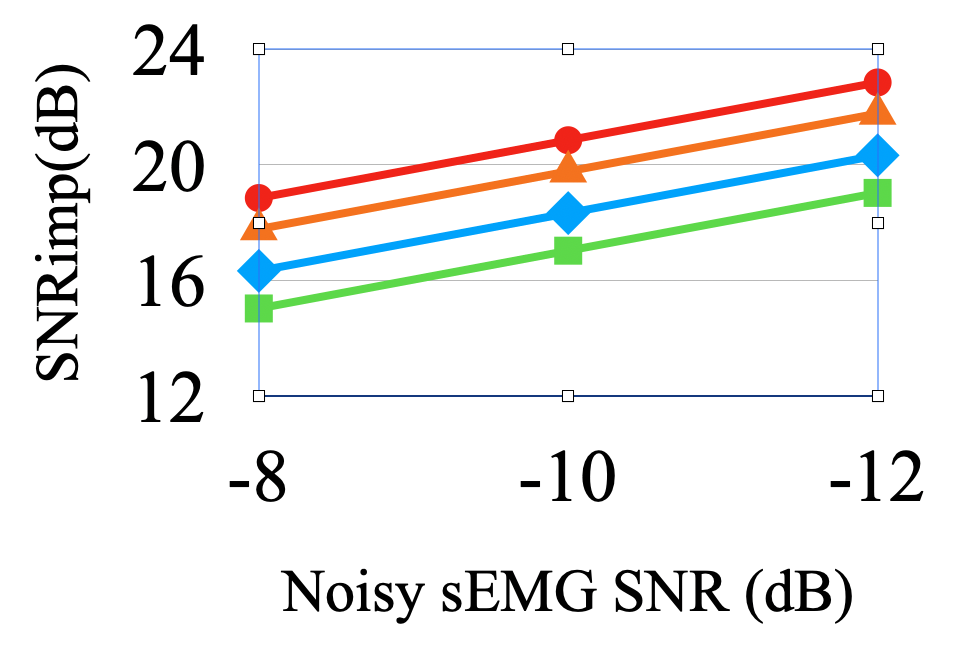}}
  \smallskip
\end{minipage}
\hfill
\begin{minipage}[b]{0.48\linewidth}
  \centering
  \centerline{(b)}
  \centerline{\includegraphics[width=4.3cm]{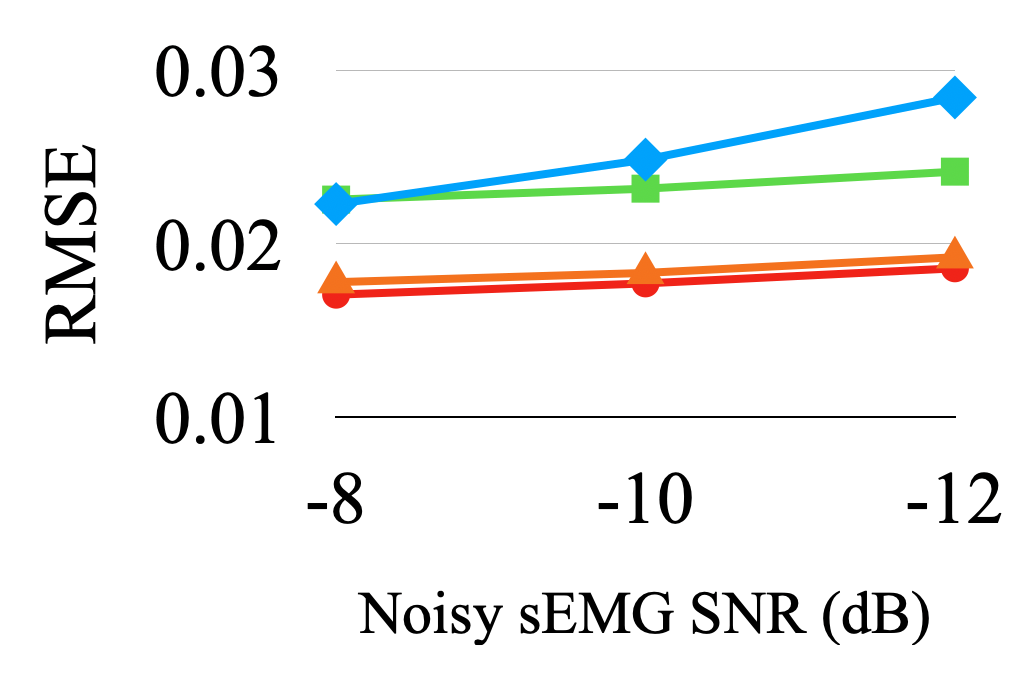}}
\end{minipage}
\begin{minipage}[b]{.48\linewidth}
  \centering
  \centerline{(c)}
  \centerline{\includegraphics[width=4.5cm]{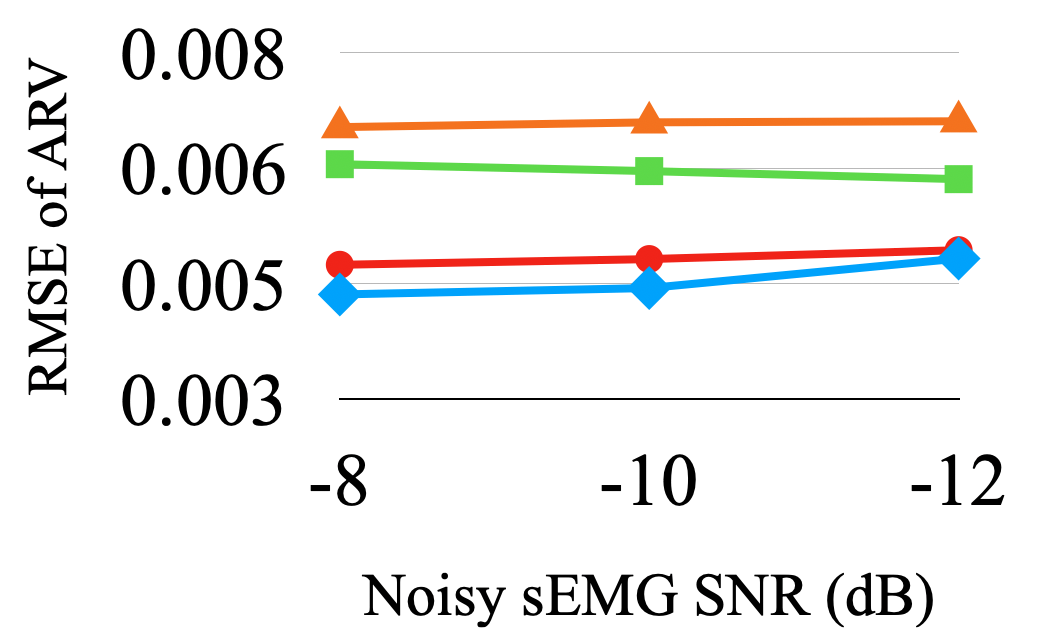}}
  \smallskip
\end{minipage}
\hfill
\begin{minipage}[b]{0.48\linewidth}
  \centering
  \centerline{(d)}
  \centerline{\includegraphics[width=4.0cm]{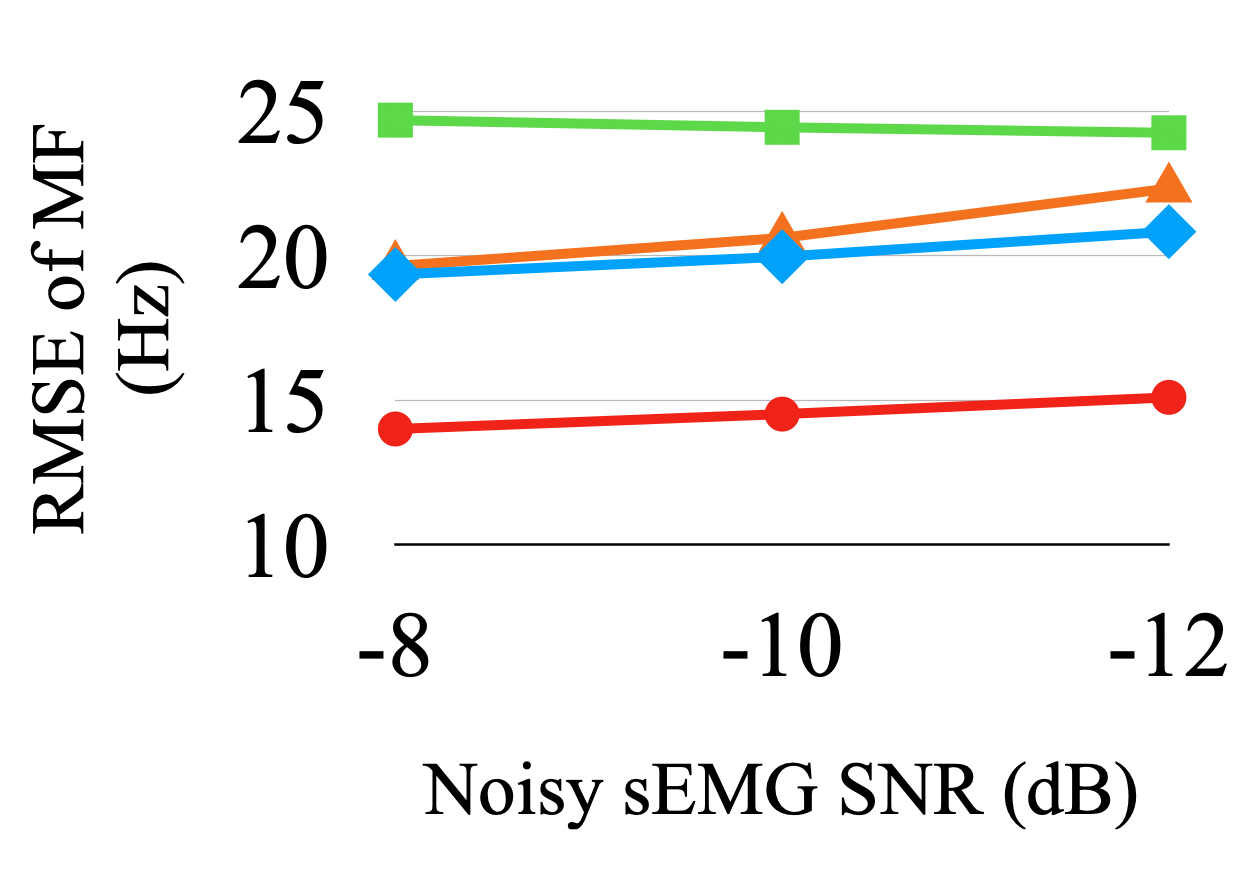}}
  
\end{minipage}
\caption{Performance of simulated trunk sEMG denoising results evaluated by (a) SNR$_{imp}$, (b) RMSE, (c) RMSE$_{ARV}$, and (d) RMSE$_{MF}$}
\label{fig: all metrics}
\end{figure}

Notably, this study evaluated the performance of SDEMG under the specific scenario in which biceps brachii sEMG~\cite{xu2020comparative,drake2006elimination} (Channel 11 in DB2) is introduced as simulation data. These trunk sEMG signals are prone to contamination of ECG at around SNR -10dB according to a previous study~\cite{zhou2006eliminating}. Consequently, we investigate the performance of SDEMG in this experimental setup along with other methods, as shown in Fig.~\ref{fig: all metrics}. The analysis demonstrates that the use of SDEMG retains its superiority in most of the cases.
Moreover, compared to our previous study~\cite{wang2023ecg}, we observed performance degradation of FCN, particularly in feature extraction error metrics. This discrepancy might stem from the reduced length of sEMG segments (from 60s to 5s), which is designed to make the denoising framework more efficient and practical. It is discovered that FCN trained with a smaller segment size tends to introduce more distortion to sEMG signals, increasing the errors in sEMG feature extraction. Conversely, SDEMG demonstrates the capability to alleviate signal distortion, generating higher-quality sEMG signals. This highlights the potential of SDEMG to provide improved signal quality for clinical assessment and evaluation.

\begin{figure}[t!]
    \centering
    \includegraphics[width=\linewidth]{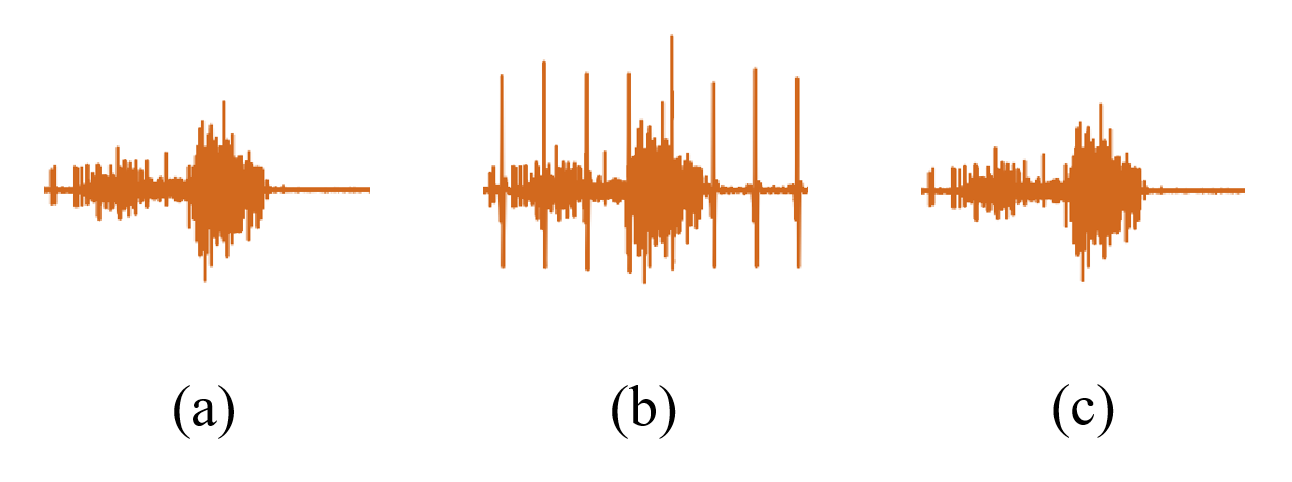}
    \caption{The waveform of (a) clean sEMG (b) noisy sEMG  (c) denoised sEMG by SDEMG.}
    \label{fig: framework}
\end{figure}

Fig.~\ref{fig: framework} presents an example of ECG contamination removal using SDEMG. It can be observed that the ECG artifacts in the noisy sEMG (SNR=-8 dB) are eliminated in the denoised waveform, and the sEMG waveform exhibits minimal distortion when compared to the clean sEMG. This underscores the capability of SDEMG to provide high-quality sEMG signals. One challenge of SDEMG is its relatively high computational effort for optimal performance. This issue may be addressed by involving ODE solvers or applying parameters pruning and quantization.

\begin{figure}[t!]
\begin{minipage}[b]{1\linewidth}
  \centering
  \centerline{\includegraphics[width=4.0cm]{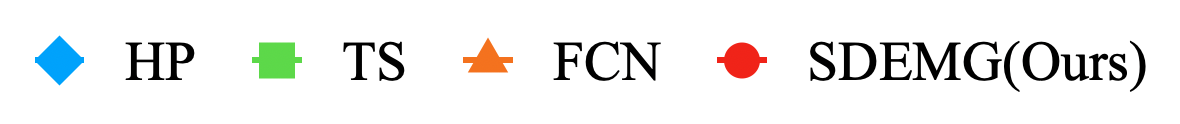}}
\end{minipage}
\end{figure}

\section{Conclusion}
\label{sec:conclusion}
In this study, we have proposed SDEMG, a score-based diffusion model, to reconstruct high-quality and high-fidelity sEMG samples from ECG-interfered sEMG signals. SDEMG requires no additional reference signal and can directly process raw sEMG signal. The experimental results demonstrate that the proposed approach outperforms the comparative methods across a broad range of SNRs and various evaluation metrics under training-testing mismatched conditions. To the best of our knowledge, this is the first study developing diffusion models for sEMG denoising. In the future, we will explore applying SDEMG on a real contaminated sEMG dataset. Moreover, we will further investigate the ability of SDEMG to perform denoising on other kinds of sEMG contaminants.


\vfill\pagebreak

\bibliographystyle{IEEEbib}
\footnotesize
\bibliography{refs}

\begin{thebibliography}{10}

\bibitem{tang2018novel}
Xiao Tang, Xu~Zhang, Xiaoping Gao, et~al.,
\newblock ``A novel interpretation of sample entropy in surface
  electromyographic examination of complex neuromuscular alternations in
  subacute and chronic stroke,''
\newblock {\em IEEE Transactions on Neural Systems and Rehabilitation
  Engineering}, vol. 26, no. 9, pp. 1878--1888, 2018.

\bibitem{engdahl2015surveying}
Susannah~M Engdahl, Breanne~P Christie, et~al.,
\newblock ``Surveying the interest of individuals with upper limb loss in novel
  prosthetic control techniques,''
\newblock {\em Journal of neuroengineering and rehabilitation}, vol. 12, no. 1,
  pp. 1--11, 2015.

\bibitem{wijsman2013wearable}
Jacqueline Wijsman, Bernard Grundlehner, Hao Liu, et~al.,
\newblock ``Wearable physiological sensors reflect mental stress state in
  office-like situations,''
\newblock in {\em 2013 humaine association conference on affective computing
  and intelligent interaction}. IEEE, 2013, pp. 600--605.

\bibitem{domnik2020clinical}
Nicolle~J Domnik, Emil~S Walsted, and Daniel Langer,
\newblock ``Clinical utility of measuring inspiratory neural drive during
  cardiopulmonary exercise testing (cpet),''
\newblock {\em Frontiers in Medicine}, vol. 7, pp. 483, 2020.

\bibitem{vandenbussche2015assessment}
Nele~L Vandenbussche, Sebastiaan Overeem, Johannes~P van Dijk, Pieter~Jan
  Simons, and Dirk~A Pevernagie,
\newblock ``Assessment of respiratory effort during sleep: esophageal pressure
  versus noninvasive monitoring techniques,''
\newblock {\em Sleep medicine reviews}, vol. 24, pp. 28--36, 2015.

\bibitem{ma2014hand}
Jiaxin Ma, Nitish~V Thakor, and Fumitoshi Matsuno,
\newblock ``Hand and wrist movement control of myoelectric prosthesis based on
  synergy,''
\newblock {\em IEEE Transactions on Human-Machine Systems}, vol. 45, no. 1, pp.
  74--83, 2014.

\bibitem{xu2020comparative}
Lin Xu, Elisabetta Peri, Rik Vullings, et~al.,
\newblock ``Comparative review of the algorithms for removal of
  electrocardiographic interference from trunk electromyography,''
\newblock {\em Sensors}, vol. 20, no. 17, pp. 4890, 2020.

\bibitem{guo2023morphological}
Liang Guo, Zhi-Wei Li, Han Zhang, Shuang-Miao Li, and Jian-Heng Zhang,
\newblock ``Morphological ecg subtraction method for removing ecg artifacts
  from diaphragm emg,''
\newblock {\em Technology and Health Care}, , no. Preprint, pp. 1--13, 2023.

\bibitem{winter2009biomechanics}
David~A Winter,
\newblock {\em Biomechanics and motor control of human movement},
\newblock John Wiley \& Sons, 2009.

\bibitem{drake2006elimination}
Janessa~DM Drake and Jack~P Callaghan,
\newblock ``Elimination of electrocardiogram contamination from electromyogram
  signals: An evaluation of currently used removal techniques,''
\newblock {\em Journal of electromyography and kinesiology}, vol. 16, no. 2,
  pp. 175--187, 2006.

\bibitem{lu2013speech}
Xugang Lu, Yu~Tsao, Shigeki Matsuda, and Chiori Hori,
\newblock ``Speech enhancement based on deep denoising autoencoder.,''
\newblock in {\em Interspeech}, 2013, vol. 2013, pp. 436--440.

\bibitem{chiang2019noise}
Hsin-Tien Chiang, Yi-Yen Hsieh, Szu-Wei Fu, et~al.,
\newblock ``Noise reduction in ecg signals using fully convolutional denoising
  autoencoders,''
\newblock {\em Ieee Access}, vol. 7, pp. 60806--60813, 2019.

\bibitem{kale2009intelligent}
Sujata~N Kale and Sanjay~V Dudul,
\newblock ``Intelligent noise removal from emg signal using focused time-lagged
  recurrent neural network,''
\newblock {\em Applied Computational Intelligence and Soft Computing}, vol.
  2009, 2009.

\bibitem{wang2023ecg}
Kuan-Chen Wang, Kai-Chun Liu, Sheng-Yu Peng, and Yu~Tsao,
\newblock ``Ecg artifact removal from single-channel surface emg using fully
  convolutional networks,''
\newblock in {\em ICASSP 2023-2023 IEEE International Conference on Acoustics,
  Speech and Signal Processing (ICASSP)}. IEEE, 2023, pp. 1--5.

\bibitem{ho2020denoising}
Jonathan Ho, Ajay Jain, and Pieter Abbeel,
\newblock ``Denoising diffusion probabilistic models,''
\newblock {\em Advances in neural information processing systems}, vol. 33, pp.
  6840--6851, 2020.

\bibitem{song2020score}
Yang Song, Jascha Sohl-Dickstein, Diederik~P Kingma, Abhishek Kumar, Stefano
  Ermon, and Ben Poole,
\newblock ``Score-based generative modeling through stochastic differential
  equations,''
\newblock {\em arXiv preprint arXiv:2011.13456}, 2020.

\bibitem{chen2020wavegrad}
Nanxin Chen, Yu~Zhang, Heiga Zen, Ron~J Weiss, Mohammad Norouzi, and William
  Chan,
\newblock ``Wavegrad: Estimating gradients for waveform generation,''
\newblock {\em arXiv preprint arXiv:2009.00713}, 2020.

\bibitem{lu2021study}
Yen-Ju Lu, Yu~Tsao, and Shinji Watanabe,
\newblock ``A study on speech enhancement based on diffusion probabilistic
  model,''
\newblock in {\em 2021 Asia-Pacific Signal and Information Processing
  Association Annual Summit and Conference (APSIPA ASC)}. IEEE, 2021, pp.
  659--666.

\bibitem{lu2022conditional}
Yen-Ju Lu, Zhong-Qiu Wang, Shinji Watanabe, Alexander Richard, Cheng Yu, and
  Yu~Tsao,
\newblock ``Conditional diffusion probabilistic model for speech enhancement,''
\newblock in {\em ICASSP 2022-2022 IEEE International Conference on Acoustics,
  Speech and Signal Processing (ICASSP)}. IEEE, 2022, pp. 7402--7406.

\bibitem{li2023descod}
Huayu Li, Gregory Ditzler, Janet Roveda, and Ao~Li,
\newblock ``Descod-ecg: Deep score-based diffusion model for ecg baseline
  wander and noise removal,''
\newblock {\em IEEE Journal of Biomedical and Health Informatics}, 2023.

\bibitem{junior2019template}
Jos{\'e} Dilermando~Costa Junior, Jos{\'e}~Manoel de~Seixas, et~al.,
\newblock ``A template subtraction method for reducing electrocardiographic
  artifacts in emg signals of low intensity,''
\newblock {\em Biomedical Signal Processing and Control}, vol. 47, pp.
  380--386, 2019.

\bibitem{kong2020diffwave}
Zhifeng Kong, Wei Ping, Jiaji Huang, Kexin Zhao, and Bryan Catanzaro,
\newblock ``Diffwave: A versatile diffusion model for audio synthesis,''
\newblock {\em arXiv preprint arXiv:2009.09761}, 2020.

\bibitem{huang2023noise2music}
Qingqing Huang, Daniel~S Park, Tao Wang, Timo~I Denk, Andy Ly, Nanxin Chen,
  Zhengdong Zhang, Zhishuai Zhang, Jiahui Yu, Christian Frank, et~al.,
\newblock ``Noise2music: Text-conditioned music generation with diffusion
  models,''
\newblock {\em arXiv preprint arXiv:2302.03917}, 2023.

\bibitem{hyvarinen2005estimation}
Aapo Hyv{\"a}rinen and Peter Dayan,
\newblock ``Estimation of non-normalized statistical models by score
  matching.,''
\newblock {\em Journal of Machine Learning Research}, vol. 6, no. 4, 2005.

\bibitem{song2020sliced}
Yang Song, Sahaj Garg, Jiaxin Shi, and Stefano Ermon,
\newblock ``Sliced score matching: A scalable approach to density and score
  estimation,''
\newblock in {\em Uncertainty in Artificial Intelligence}. PMLR, 2020, pp.
  574--584.

\bibitem{song2019generative}
Yang Song and Stefano Ermon,
\newblock ``Generative modeling by estimating gradients of the data
  distribution,''
\newblock {\em Advances in neural information processing systems}, vol. 32,
  2019.

\bibitem{nichol2021improved}
Alexander~Quinn Nichol and Prafulla Dhariwal,
\newblock ``Improved denoising diffusion probabilistic models,''
\newblock in {\em International Conference on Machine Learning}. PMLR, 2021,
  pp. 8162--8171.

\bibitem{romero2021deepfilter}
Francisco~P Romero, David~C Pi{\~n}ol, and Carlos~R V{\'a}zquez-Seisdedos,
\newblock ``Deepfilter: An ecg baseline wander removal filter using deep
  learning techniques,''
\newblock {\em Biomedical Signal Processing and Control}, vol. 70, pp. 102992,
  2021.

\bibitem{chen2021hinet}
Liangyu Chen, Xin Lu, Jie Zhang, Xiaojie Chu, and Chengpeng Chen,
\newblock ``Hinet: Half instance normalization network for image restoration,''
\newblock in {\em Proceedings of the IEEE/CVF Conference on Computer Vision and
  Pattern Recognition}, 2021, pp. 182--192.

\bibitem{dumoulin2018feature}
Vincent Dumoulin, Ethan Perez, Nathan Schucher, Florian Strub, Harm~de Vries,
  Aaron Courville, and Yoshua Bengio,
\newblock ``Feature-wise transformations,''
\newblock {\em Distill}, vol. 3, no. 7, pp. e11, 2018.

\bibitem{atzori2014electromyography}
Manfredo Atzori, Arjan Gijsberts, Claudio Castellini, et~al.,
\newblock ``Electromyography data for non-invasive naturally-controlled robotic
  hand prostheses,''
\newblock {\em Scientific data}, vol. 1, no. 1, pp. 1--13, 2014.

\bibitem{machado2021deep}
Juliano Machado, Amauri Machado, and Alexandre Balbinot,
\newblock ``Deep learning for surface electromyography artifact contamination
  type detection,''
\newblock {\em Biomedical Signal Processing and Control}, vol. 68, pp. 102752,
  2021.

\bibitem{goldberger2000physiobank}
Ary~L Goldberger, Luis~AN Amaral, Leon Glass, et~al.,
\newblock ``Physiobank, physiotoolkit, and physionet: components of a new
  research resource for complex physiologic signals,''
\newblock {\em circulation}, vol. 101, no. 23, pp. e215--e220, 2000.

\bibitem{zhang2021eegdenoisenet}
Haoming Zhang, Mingqi Zhao, Chen Wei, et~al.,
\newblock ``Eegdenoisenet: A benchmark dataset for deep learning solutions of
  eeg denoising,''
\newblock {\em Journal of Neural Engineering}, vol. 18, no. 5, pp. 056057,
  2021.

\bibitem{zhou2006eliminating}
Ping Zhou and Todd~A Kuiken,
\newblock ``Eliminating cardiac contamination from myoelectric control signals
  developed by targeted muscle reinnervation,''
\newblock {\em Physiological Measurement}, vol. 27, no. 12, pp. 1311, 2006.

\end{thebibliography}

\end{document}